\documentclass[conference,letterpaper]{IEEEtran}

\usepackage[utf8]{inputenc} 
\usepackage[T1]{fontenc}
\usepackage[cmex10]{amsmath} % Use the [cmex10] 
\usepackage{epsfig,rotating,setspace,latexsym,amsmath,epsf,amssymb,amsfonts,bm,theorem,cite,enumerate,longtable,accents,float,physics}
\usepackage{algorithm,graphicx,epsf,authblk,epstopdf,url,xcolor, soul,multirow,bbm}
\usepackage{mathtools,comment}
\usepackage[center]{qtree}
\usepackage{tree-dvips}
\usepackage[noend]{algpseudocode}
\usepackage[linguistics]{forest }

\usepackage{dsfont}

\newtheorem{theorem}{Theorem}

\newtheorem{definition}{Definition}
\newtheorem{remark}{Remark}
\newtheorem{lemma}{Lemma}

\DeclareMathOperator{\diag}{diag}
\title{Private Counterfactual Retrieval}

\author{Mohamed Nomeir \quad Pasan Dissanayake  \quad Shreya Meel \quad Sanghamitra Dutta \quad Sennur Ulukus \\
\normalsize Department of Electrical and Computer Engineering \\
\normalsize University of Maryland, College Park, MD 20742 \\
\normalsize {\it mnomeir@umd.edu} \quad {\it pasand@umd.edu} \quad{\it smeel@umd.edu} \quad {\it sanghamd@umd.edu} \quad {\it ulukus@umd.edu}}

\begin{document}

\maketitle

\begin{abstract}
Transparency and explainability are two extremely important aspects to be considered when employing black-box machine learning models in high-stake applications. Providing counterfactual explanations is one way of fulfilling this requirement. However, this also poses a threat to the privacy of both the institution that is providing the explanation as well as the user who is requesting it. In this work, we propose multiple schemes inspired by private information retrieval (PIR) techniques which ensure the \emph{user's privacy} when retrieving counterfactual explanations. We present a scheme which retrieves the \emph{exact} nearest neighbor counterfactual explanation from a database of accepted points while achieving perfect (information-theoretic) privacy for the user. While the scheme achieves perfect privacy for the user, some leakage on the database is inevitable which we quantify using a mutual information based metric. Furthermore, we propose strategies to reduce this leakage to achieve an advanced degree of database privacy. We extend these schemes to incorporate user's preference on transforming their attributes, so that a more actionable explanation can be received. Since our schemes rely on finite field arithmetic, we empirically validate our schemes on real datasets to understand the trade-off between the accuracy and the finite field sizes. Finally, we present numerical results to support our theoretical findings, and compare the database leakage of the proposed schemes.
\end{abstract}


\section{INTRODUCTION}

%Para 1: Counterfactual explanations are generating immense interest 
With the growing call for the right to an explanation~\cite{voigt2017eu,park2023ai}, the framework of counterfactual explanations has generated immense interest as a means to explain the decision-making of complex models in high-stakes applications~\cite{Harvard_discussion_first_counterfactual}. Counterfactual explanations provide the minimum input perturbation required to alter a model outcome, and are closely tethered to algorithmic recourse, e.g., increase your income by 10K to qualify for a loan. 

Counterfactual explanations are also susceptible to privacy concerns. For instance, \cite{pawelczykMembershipInference,yang2022differentially} bring out privacy issues related to the underlying training data, while \cite{aivodjiModelExtraction,wangDualCFModelExtraction,dissanayakeModelExtraction} study model extraction using counterfactual explanations. However, these works predominantly focus on privacy from the institution's side. Applicant privacy concern arises if they wish to obtain their counterfactual explanations \emph{privately} without revealing their current input feature vector to the institution.

%Para 2: Use cases of pcr
An applicant may be reluctant to share their entire feature vector with the institution for several reasons, e.g., a formal application process might be expensive in terms of time and resources, or allow for only a limited number of attempts~\cite{dissanayakeModelExtraction}, or they might wish to preserve the privacy of their data until they improve their chances of acceptance.  

%Para 3:
Our work introduces the novel problem of \textsl{private counterfactual retrieval (PCR)}. The objective of PCR is to design a strategy that applicants and institutions can jointly agree upon that enables: (i) the applicant to \textsl{privately} retrieve their counterfactual explanation from an institution through an alternate set of queries; and (ii) the institution to also not leak any further information beyond what the applicant requires. To this end, we draw inspiration from the problem of private information retrieval (PIR)~\cite{chor,ulukusPIRLC} which enables a user to download a message from a set of messages stored in a system of databases without revealing the index of the desired message. Along the lines of PIR, we assume that the institution has a stored database of accepted applicants, and the database entries lie in a finite field. We seek to retrieve the index of the \textsl{exact} nearest neighbor for an applicant from the database without revealing their own feature vector (in an information-theoretic sense).  This yields a new and complementary formulation and renders our work incomparable to the prior works since they do not attain perfect information-theoretic privacy.

Notably, the key difference between PIR and PCR is as follows: In PIR, a user knows the index of the required message, whereas in PCR the user does not know the index of the sample they will retrieve, except that the sample is closest, in some sense, to their own feature vector, posing additional challenges. Our work proposes novel strategies for PCR that enable the applicant to achieve perfect privacy while limiting the leakage from the institution's side. PCR could also be viewed as a novel (and nontrivial) version of the PIR problem which could also be of independent interest outside the counterfactual context.

To summarize, our main contributions are:
 \begin{enumerate}
 % [topsep=-1pt,itemsep=0ex,partopsep=-1ex,parsep=1ex, leftmargin=*]
    \item We introduce the novel problem of \textbf{private counterfactual retrieval (PCR)}, along with a baseline scheme to achieve user privacy to retrieve the index of the closest counterfactual using the $\ell_2$ distance metric.
    \item We develop two different PCR schemes that we call Diff-PCR and Mask-PCR to provide the institution with better privacy for their database compared to the baseline scheme while achieving perfect information-theoretic privacy for the applicant.
    \item We also incorporate actionability for the applicant as an additional criterion in our design, proposing an extended scheme that we call PCR+.
    \item We also perform an empirical analysis to understand the implications of our finite field assumptions on real data. We compare the accuracy loss from translating real-valued data to finite-field data to ensure that the designed schemes act as intended, and to understand the trade-off between the field size requirements and the efficacy of our schemes.
    \item Finally, we numerically evaluate the database leakage for the proposed PCR schemes, and verify the comparative leakage values among the proposed schemes.
\end{enumerate}

\subsection{RELATED WORKS}
\paragraph{Counterfactual explanations and privacy} Since the initial formulation in \cite{Harvard_discussion_first_counterfactual}, numerous works have been focusing on generating counterfactual explanations with different properties. Proximity to the query instance \cite{nice}, robustness \cite{upadhyayROAR, hammanRobustCF}, actionability \cite{face, dice}, sparsity in change \cite{nice, dice}, and diversity \cite{dice} are some such properties; we refer the reader to \cite{karimiCFMethodsSurvey} and \cite{guidottiCounterfactualSurvey} for a comprehensive survey on different methods. In this work, we focus on both proximity to the original instance and actionability. Moreover, we use nearest-neighbor counterfactuals as the explanation method considered. Existing works on privacy within the context of counterfactual explanations mainly focus on the institution's end. In this regard, \cite{pawelczykMembershipInference} analyzes inferring the membership of an explanation in the training set of the model while \cite{yang2022differentially} provides differentially-private counterfactuals. \cite{aivodjiModelExtraction} and \cite{wangDualCFModelExtraction} present two ways of utilizing counterfactuals to extract the model when counterfactuals are provided for any query. \cite{dissanayakeModelExtraction} presents a model extraction strategy that specifically leverages the fact that the counterfactuals are close to the decision boundary. Explanation linkage attacks that try to extract private attributes of a nearest-neighbor counterfactual explanation are discussed in \cite{privacy_issue_in_cf}. All of these works focus on the privacy of either the model or the data stored in the institution's database. In contrast, we are interested in the privacy of the applicant who is asking for an explanation. 

\paragraph{Private information retrieval}
The PIR problem formulation was first defined in \cite{chor} and its capacity, i.e., the maximum ratio between the number of required message symbols and the number of total downloaded symbols, was found in \cite{c_pir}. Different from the original formulation which required non-colluding databases with replicated contents, other variants, such as PIR with colluding databases \cite{colluding, arbitrarycollusion}, PIR with coded databases, \cite{Salim_CodedPIR, banawan_pir_mdscoded} were also considered. In SPIR, an extra requirement is that the user cannot get any information beyond its required message. The capacity of SPIR was found in \cite{c_spir} and the variants with colluding and coded databases appeared in \cite{tspir_mdscoded}. Reference \cite{csa}, proposes a cross-subspace alignment (CSA) approach as a unifying framework for PIR and SPIR with additional requirements, such as security against the storing units, e.g., servers. These schemes are capacity-achieving in some cases, for instance, when the number of messages is large. We refer the reader to \cite{ulukusPIRLC} for a comprehensive survey on the PIR and SPIR literature. 

\textbf{Private approximate nearest neighbor search:} Closely related to our problem is the nearest neighbor search problem, where the user needs to retrieve the indices of the vectors in a database, that are closest to their vector according to some similarity metric. In this regard, \cite{ANN_sublinear} proposed algorithms that guarantee computational privacy, both to the user and the database, while the user retrieves a sub-optimal nearest neighbor. Reference \cite{sajani_ANN} proposed an information-theoretically private clustering-based solution based on the dot-product metric. This work, however, did not consider database privacy, and the user retrieves only an approximate nearest neighbor.

\section{SYSTEM MODEL}
\begin{figure}[t]
\centering
\includegraphics[width=0.25\textwidth ]{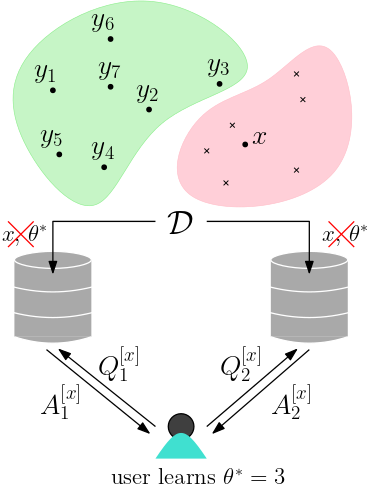}
\caption{System model with $w=\mathbf{1}^t$. The green and red regions represent the accepted and rejected samples respectively. The user learns that $\theta^*=3$ is their counterfactual index in $\mathcal{D}$.}
\label{system_model_case1}
\end{figure}

The institution has a pre-trained binary classification model that takes as input a $d$-dimensional feature vector and classifies it into its target class, e.g., accepted or rejected. A user who is rejected by this model wishes to privately retrieve a valid counterfactual sample corresponding to their data sample. However, the user does not have access to the model and relies on a database $\mathcal{D}$ that contains the feature vectors of a set of samples accepted by the  model as depicted in Fig.~\ref{system_model_case1}. We assume that each attribute of the samples is an integer in $[0:R]$. The samples in $\mathcal{D}$ are stored in $N$ non-colluding and non-communicating servers in a replicated manner. The samples are indexed as $y_1, y_2, \ldots, y_M$ where $M=|\mathcal{D}|$. The goal of the user is to retrieve the index of the accepted data sample that is nearest to their sample $x$ based on a preference vector\footnote{We consider the system model based on actionable vector $w$ as the general system model. $w$ which is globally known. Our system can be simplified to the case of equal actionability on all features by choosing $w = \mathbf{1}^t$.}, $w$, according to some distance metric $d_w(.,.)$, i.e.,
\begin{align}
    \theta^*=\arg\!\min_{i: {y_i}\in \mathcal{D}} d_w(x,y_i).
\end{align}
To this end, the user sends the query $Q_n^{[x,w]}$ to server $n\in [N]$. Upon receiving the queries, each server computes their answers, $A_n^{[x,w]}$ using their storage, their queries and shared common randomness, $Z'$, i.e.,
\begin{align}
    H(A_n^{[x,w]}|\mathcal{D},Q_n^{[x,w]},Z') = 0.
\end{align}
 Using the responses from all the servers, the user determines the index $\theta^*$ of their corresponding counterfactual, i.e., 
\begin{align}
    [\text{Decodability}] \quad H(\theta^*| Q_{[N]}^{[x,w]}, A_{[N]}^{[x,w]}, x,w) = 0.
\end{align}
Since user privacy is the main concern, each server must not acquire any information on the user's sample or the index of their counterfactual, i.e., $\forall n \in [N]$
\begin{align}
   [\text{User Privacy}] \quad I(x,w, \theta^*; Q_n^{[x,w]}, A_n^{[x,w]}| \mathcal{D})=0.
\end{align}
To quantify privacy for the servers, we consider an information-theoretic measure that defines the amount of information leakage about the samples in $\mathcal{D}$ to the user, upon receiving the answers, as follows,
\begin{align}\label{eq:leakage_pcr}
[\text{Database Leakage}] \quad I(y_1,\ldots,y_M; Q_{[N]}^{[x,w]}, A_{[N]}^{[x,w]}|x,w)
\end{align}
 Our goal is to develop schemes that reduce this database leakage, while maintaining perfect user privacy. 
% The requirements stated above can be formulated as an optimization problem as follows 
% \begin{align}\label{opt_1}
%     \min_{\mathcal{A}, \mathcal{Q}, \mathcal{S}}\quad &I(y_1,\ldots,y_M; Q_{[N]}^{[x,w]}, A_{[N]}^{[x,w]}|x,w)\\
%     \text{s.t. }    &H(\theta^*|A_{[N]}^{[x,w]},x,w)=0\\
%     & I(x,w, \theta^*; Q_n^{[x,w]}, A_n^{[x,w]}| y_1, \ldots, y_M)=0
% \end{align}
% where $\mathcal{A}$ is the set of all possible answers the servers can send, $\mathcal{Q}$ are all the possible queries the user can send, and $\mathcal{S}$ are all the possible ways to encode and store the accepted samples in the $N$ servers. 
We use the (weighted) $\ell_2$ norm, i.e., $d_w(u,v)=(u-v)^T \bm{W} (u-v)$ where $\bm{W}$ is the diagonal matrix with $w$ on its main diagonal. Our results can be extended easily using any $\ell_n$ norm with $n$ being even, as well as to the dot product metric\footnote{It is important to note that if dot product metric is used, actionability can be implemented directly in the rejected sample sent by the user.}. For space limitations, we define Vandermonde matrices on the finite field $\mathbb{F}_q$ where $q$ is a positive prime power.

The benefit of privately retrieving the index of the closest neighbor is to guarantee that the institution does not learn any information about the user’s input. Further, it does not learn the counterfactual explanation corresponding to the user. This is achieved in two steps: First, we perform the PCR scheme to retrieve the counterfactual index from the server. Next, we use this index to perform the SPIR scheme \cite{c_spir} on the same dataset. The SPIR scheme applies seamlessly in the existing system due to the replicated and non-colluding databases.

\begin{definition}[Vandermonde matrices]
    Let $\alpha_1,\ldots,\alpha_n$ be distinct elements of $\mathbb{F}_q$. Let  $\bm{M}_n$ denote the $n\times n$ Vandermonde matrix. Then, 
    \begin{align}
        \bm{M}_n(i,j) = \alpha_i^{j-1}, \quad i,j \in [n].
    \end{align}
    Moreover, $\bm{M}_n$ is invertible in $\mathbb{F}_q$.
\end{definition}

% In addition, for completeness, the one-time pad theorem, which we use extensively, is stated below as in \cite{coverthomas}.
% \begin{theorem}[One-time pad theorem]
%     Let $q$ be a prime, then for any random variable $X \in \mathbb{F}_q$, and any uniform random variable $Z$ chosen at random in $\mathbb{F}_q$
%     \begin{align}
%         I(X; X+Z)=0,
%     \end{align}
%     where addition is according to $\mathbb{F}_q$ arithmetic.
% \end{theorem}

We divide the description of our schemes into two sections, Section \ref{no_pref} without user preference and Section \ref{private_pref} with private user actionability.

Finally, it is important to note that this formulation for counterfactual retrieval is novel, as far as we know. This is mainly because perfect user privacy is required and the servers accept some data leakage. Thus it is unfair to compare our model and schemes developed here with other counterfactual retrieval schemes.

\section{PROPOSED PCR SCHEMES}\label{no_pref}
In this section, we assume $w=\mathbf{1}^t$, thus $d_{w}(.,.)$ is the squared $\ell_2$ norm. Consequently, we simplify the notations of queries and answers for server $n$ to $Q_n^{[x]}$ and $A_n^{[x]}$, respectively. For ease of exposition, we denote
\begin{align}
    d_i(x) =  ||y_i-x||^2 , \quad \forall i\in [M].
\end{align}
First, we present the \emph{baseline} scheme taking only user privacy into consideration. Then, we present two approaches, Diff-PCR and Mask-PCR that leak less information on $\mathcal{D}$ than the baseline, in addition to achieving user privacy.

\subsection{Baseline PCR}\label{userpriv1}
\begin{theorem}\label{PCR_baseline_thm}
    There exists a scheme that can retrieve the exact closest counterfactual index from $N=2$ servers with perfect user privacy, with the communication cost of $2(d+M)$ symbols of $\mathbb{F}_q$ where $q>R^2d$ is prime.
\end{theorem}
  
To prove the achievability of Theorem~\ref{PCR_baseline_thm}, we present our baseline PCR scheme. Let the operating field be $\mathbb{F}_q$ where $q$ is a prime satisfying $q>R^2d$, where $d$ is the number of features for each accepted sample and $R$ is the maximum value each feature can have. In addition, we require $q$ to be greater than the aforementioned value to make sure that addition in this field will not reverse the order of the closeness of the samples with the user sample. Although fields can be defined using prime powers, we restrict ourselves to only primes since prime power fields, $\mathbb{F}_{q^r}$, are isomorphic to $\mathbb{F}_q^r$ which means that they are not ordered. Each server stores the $M$ accepted samples, $y_1, \ldots, y_M$, each being a $d$ dimensional vector. Let $\alpha_1, \alpha_2$ be two distinct elements of $\mathbb{F}_q$ globally known to the user and servers. The user privately generates a random vector $Z$ uniformly from $\mathbb{F}_q^d$ and then sends the query $Q_n^{[x]}$ to the $n$th server based on $x$ as follows,
\begin{align}\label{queries_no_weight}
    Q_n^{[x]} = x+\alpha_n Z,
\end{align}
where $\alpha_1, \alpha_2$ are distinct elements in the field and globally known constants. Since $x$ is one-time padded with $\alpha_nZ$ which is uniform on $\mathbb{F}_q$, each server learns no information on $x$, i.e., $I(x;Q_n^{[x]}|y_1,\ldots,y_M) = 0$. Let the servers share $M$ independent random variables $Z'(i), i \in [M]$ picked uniformly from $\mathbb{F}_q$. Given a query, the servers compute one answer for each $y_i\in \mathcal{D}$ as follows,
\begin{align}
    A_n^{[x]} (i) =& ||y_i-Q_n^{[x]}||^2 + \alpha_n Z'(i) \nonumber \\
    =& d_i(x)+  \alpha_n \left(2(y_i-x)^t Z + Z'(i)\right) \nonumber \\
    &+ \alpha_n^2 ||Z||^2.
\end{align}
Note that $A_n^{[x]} (i)$ is a second degree polynomial of $\alpha_n$ with $\alpha_n^2 ||Z||^2$ already known to the user. Therefore, the user cancels $\alpha_1^2||Z||^2$ and $\alpha_2^2||Z||^2$ from $A_1^{[x]}(i)$ and $A_2^{[x]}(i)$, respectively, 
\begin{align}\label{get_d_i(x)}
    \!\!\!\begin{bmatrix}
        A_1^{[x]}(i)-\alpha_1^2||Z||^2\\
        A_2^{[x]}(i)-\alpha_2^2||Z||^2
    \end{bmatrix}=
   \bm{M}_2
    \begin{bmatrix}
        d_i(x)\\
        2 (y_i-x)^t Z + Z'(i)
    \end{bmatrix}\!. \!\!\!
\end{align}
The user compares the values of $ d_i(x)\in \mathbb{F}_q$, for all $i\in [M]$ and assigns $\theta^*$ to the $i$ for which this is minimum. 

\paragraph{Communication Cost} This scheme requires two $d$-dimensional vectors of $\mathbb{F}_q$ to be sent, one to each server. This entails an upload cost of $2d$ symbols. Each server responds with $M$ symbols from $\mathbb{F}_q$, thereby, incurring a download cost of $2M$.

\paragraph{Computation Complexity} {At the user side, the computation for the queries involves two scalar multiplications and additions in $d$ dimensions, which is $O(d)$. To find $\theta^*$, the user decodes $d_i(x)$ using \eqref{get_d_i(x)} for each $i\in [M]$ and compares them to find the minimum, both using $O(M)$ operations. Therefore, the user-side computational complexity is $O(d+M)$. Each server generates $M$ answers, where each answer involves subtraction, squaring, and addition of $d$ elements of each vector. This results in $O(Md)$ operations at each server.}

\begin{remark}
    It is important to note that \eqref{remark_edit} is evaluated prior to the scheme initiation and done once only till the dataset itself changes. This is a part of pre-processing for Mask-PCR and is not a part of the main scheme, which begins with the user sending their query. This is why it is not included in the calculation of the computation complexity. 
\end{remark}

\subsection{Diff-PCR}\label{difference_distance_equal_actionable}
\begin{theorem}\label{PCR_diff_thm}
    There exists a scheme  with a lower value of database leakage \eqref{eq:leakage_pcr} compared to the baseline PCR, with the communication cost of $2(d+M-1)$ symbols of $\mathbb{F}_q$ where $q>2R^2d$ is prime.
\end{theorem}

We show that we can improve database privacy by revealing only the difference of norms while maintaining the user's privacy. We show that this can be accomplished with $N=2$ replicated databases as in the baseline PCR. The field of operation is a prime $q> 2R^2d$. This is required because for all $i,j\in [M]$,
\begin{align}
    0\leq| d_i(x)- d_j(x)|\leq R^2d.
\end{align}
Therefore, 
\begin{align}
    &d_i(x)- d_j(x)\in \! \! \! \!
    &\begin{cases} 
        [0:R^2d], & \! \! d_i(x)\geq d_j(x)\\
        [R^2d+1:q-1], & \! \!  d_i(x)<d_j(x)
    \end{cases}.
\end{align}
The servers share a common random vector $Z'_{[M-1]}=[Z'(1) , Z'(2), \ldots, Z'(M-1)]^t$, where each entry is picked uniformly and independently from $\mathbb{F}_q$. As described in Section \ref{userpriv1}, the user sends the query given in \eqref{queries_no_weight} to server $n=1,2$. Then, server $n$ computes the following answer for each $i\in [M-1]$,
\begin{align}
    A_n^{[x]}(i)=&||y_i-Q_n^{[x]}||^2 - ||y_{i+1}-Q_n^{[x]}||^2 + \alpha_n Z'(i) \nonumber\\
    =&||y_i-x||^2 - ||y_{i+1}-x||^2 +\alpha_n\left(2(y_{i+1}-x)^t Z \right. \nonumber\\
    &\left.-2(y_i-x)^t Z+ Z'(i)\right) \nonumber\\
    =& d_i(x)- d_{i+1}(x) +\alpha_n I'(i),
\end{align}
where $I'(i)=2(y_{i+1}-y_i)^t Z+ Z'(i)$. Using $A^{[x]}_1(i)$ and $A^{[x]}_2(i)$, the user exactly recovers $||y_i-x||^2 - ||y_{i+1}-x||^2$, because
\begin{align}\label{get_diff_d_i}
    \begin{bmatrix}
        A_1^{[x]}(i)\\
        A_2^{[x]}(i)
    \end{bmatrix}=
    \bm{M}_2
    \begin{bmatrix}
        d_i(x)-d_{i+1}(x)\\
        I'(i)
    \end{bmatrix}.
\end{align}
Therefore, the user recovers the $M-1$ differences along with $M-1$ interference terms. In each $I'(i)$, the one-time-padding with $Z'(i)$ makes sure that no information on $y_{i+1}-y_i$ is revealed.

{Finding $y_i$ Closest to $x$:} The user finds the index $\theta^*$ of their counterfactual using Algorithm \ref{alg_difference_distance}.

\begin{algorithm}
\caption{Algorithm to compute $\theta^*$}\label{alg_difference_distance}
 \hspace*{\algorithmicindent} \textbf{Input:} $d_i(x)-d_{i+1}(x), \quad i \in [M-1]$\\
 \hspace*{\algorithmicindent} \textbf{Output:} $\theta^*$
\begin{algorithmic}[1]
    \State {$\theta^* = 1$}
    \For {$i \in [M-1]$}
    \State {$r(i)= d_i(x) - d_{i+1}(x)$}
    \State {$d_{\theta^*}(x) - d_{i+1}(x)= \sum_{j=\theta^*}^{i}r(j)$}
    \If {$d_{\theta^*}(x) - d_{i+1}(x)\in [R^2d+1:q-1]$}\label{range_prox}
    \State {$\theta^* \leftarrow \theta^*$}
    \Else
    \State {$\theta^* \leftarrow i+1$}
    \EndIf
    \EndFor
    \State \Return $\theta^*$
\end{algorithmic}
\end{algorithm}

\paragraph{Communication Cost} The upload cost incurred in this scheme is $2d$ and the download cost is $2(M-1)$.

\paragraph{Computation Complexity} The user side query generation is $O(d)$ as in the baseline scheme. To find $\theta^*$, the user finds $d_i(x)-d_{i+1}(x)$ using \eqref{get_diff_d_i} and evaluates $\theta^*$ using Algorithm \ref{alg_difference_distance}, which has a complexity of $O(M)$. Thus, the user's computational complexity is $O(d+M)$. The server side complexity is $O(Md)$, same as the baseline.

\paragraph{Leakage Analysis} To show that Diff-PCR has lower leakage, we make the following observation
\begin{align}
    &I(r(1), r(2),\ldots, r(M-1); y_1, \ldots, y_M |x) \nonumber \\
   & \leq  I(d_1(x), \ldots, d_M(x) ;y_1,\ldots,y_M | x), \label{cmt_3}
\end{align}
which is due to the data-processing inequality. 

\subsection{Mask-PCR}\label{masking_equal_actionable}
\begin{theorem}\label{PCR_mask_thm}
    There exists a scheme that has a lower database leakage in terms of \eqref{eq:leakage_pcr} compared to the baseline PCR with the communication cost of $2(d+M)$ symbols of $\mathbb{F}_q$ where $q>R^2d$ is prime.
\end{theorem}

In this approach we need the servers to be able to have access to the rejected set, i.e., $\mathcal{D}_{r} = \{x_1, \ldots, x_K\}$ (this restriction is removed during the experimental analysis). Now, define the closure of $\mathcal{D}_{r}$ as follows
\begin{align}\label{remark_edit_2}
    \mathcal{D}_c= \mathtt{clo}(\mathcal{D}_r) = \{&x \in \mathbb{F}_q^d: | d_{i}(x) - d_{j}(x) |-\nonumber\\ 
    &|d_{i}(x_k) - d_{j}(x_k)| \geq 0, \forall i, j, k\}\setminus \mathcal{D}.
\end{align}
In addition, we define the following metrics 
\begin{align}\label{remark_edit}
   % & d_{ij} = ||x_i-y_j||^2, \\
    & d_{k} = \min_{i,j} |d_{i}(x_k) - d_{j}(x_k)|, \quad  d_{\min} = \min_{k} d_k.
\end{align}
Let $\mu$ be a random variable with support $\{0,\ldots, d_{\min}-1\}$. Now, a user who wishes to know the closest accepted sample to their rejected sample $x \in \mathcal{D}_c$ sends the query in \eqref{queries_no_weight} to the $n$th server. Upon receiving the queries, each server computes the answers as follows
\begin{align}
    A_n^{[x]}(i) = ||y_{i}-Q_n^{[x]}||^2 + \mu(i) + \alpha_n Z'(i), \quad i \in [M], 
\end{align}
where $\mu(i)$ has the same distribution as $\mu$ and $Z'(i)$ is a uniform random variable in $\mathbb{F}_q$. As the user receives the answers from the servers, they are reprocessed as follows,
\begin{align}\label{pcr_ans_reprocess}
    \hat{A}_n^{[x]}(i)=A_n^{[x]}(i) - \alpha_n^2 ||Z||^2, \quad \forall i,n,
\end{align}
and are used to decode the distances as follows,
\begin{align}
    A^{[x]}(i) = \begin{bmatrix}
        \hat{A}_1^{[x]}(i)\\
        \hat{A}_2^{[x]}(i)
    \end{bmatrix}= \bm{M}_2\begin{bmatrix}
        d_i(x) + \mu(i)\\
        I(i)+ Z'(i)
    \end{bmatrix}.
\end{align}
Upon getting the values of masked distances, $d_i(x) + \mu(i)$, $i \in [M]$, the user utilizes them to decide which is closest based on their numerical value, i.e., decode the index of the closest accepted sample. To show that the user can correctly decode with the masked distances, we need the following lemma, where its proof can be found in \cite{PCR_full_journal}.

\begin{lemma}\label{Mask-PCR-decoding-lemma}
    If $x \in \mathcal{D}_c$, then the user is able to decode the index of the closest acceptable sample\footnote{In contrast to the difference approach, the field size does not need to be expanded to consider the difference issue. The main reason is that the difference calculations here are done at the user side and the field size restriction can be dropped.}.
\end{lemma}

\begin{remark}
    Lemma~\ref{Mask-PCR-decoding-lemma} states that all the vectors in $\mathcal{D}_c$ preserve the relative distances among the accepted samples, i.e., the ordering of the distances among the points in $\mathcal{D}_c$ and accepted samples is maintained. In addition, note that the user does not know the exact value of $d_{\min}$ since they do not have any prior knowledge of the accepted samples.       
\end{remark}

{
\begin{remark}\label{rmk:closure_lift}
    We believe that the requirement $x\in \mathcal{D}_c$ is a bit strong, which is why we dropped this requirement in our first experiment (see Sec \ref{exp:accu_quant}) and we find $d_{\min}$ empirically instead. However, it is required for theoretical analysis to guarantee decodability.
\end{remark}
}

\paragraph{Communication Cost} The upload cost in this scheme is $2d$ and the download cost is $2M$.

\paragraph{Computation Complexity} The computational complexity at the user and the server is equal to that of baseline PCR.
\begin{remark}
    It is important to note that \eqref{remark_edit} is evaluated prior to the scheme initiation and done once only till the dataset itself changes. This is a part of pre-processing for Mask-PCR and is not a part of the main scheme, which begins with the user sending their query. This is why it is not included in the calculation of the computation complexity. 
\end{remark}
\paragraph{Effect of Field Size in Masking}
The field size can have a significant role in counterfactual retrieval with the masking approach. Recall that the minimum requirement for the field size is $R^2d$ as explained at the beginning of this section. Let $q_1$ be the field size satisfying $q_1> R^2 d$. If another field size $q_2$ is chosen such that $q_2 > q_1$, we can embed the samples $y_1, \ldots, y_M$ in $\mathbb{F}_{q_2}$ using a relative distance preserving transform $T: \mathbb{F}_{q_1} \rightarrow \mathbb{F}_{q_2}$ such that the following condition is satisfied $\forall k,i,j$,
\begin{align}
    |d_{i}(x_k)-d_{j}(x_k)| \leq & | \big(||x_k -T(y_i)||^2- ||x_k -T(y_j)||^2\big)|.
\end{align}
The support of the random variable used in masking $\mu$ is larger compared to when the field size is $q_1$ since $d_{\min} \leq |d_{i}(x_k)-d_{j}(x_k)|$. Thus, the estimation error of the user for the exact values of the samples can increase. In addition, note that this transform can be kept hidden from the user since it does not affect the result for ordering of samples by construction.

\paragraph{Illustrative Example}
This example demonstrates the masking approach and the field size expansion. Let the rejected samples, for example, be $\{[1,2]^t, [2,1]^t\}$, and the set of accepted samples be $\{[20,0]^t,[0,20]^t\}$. Let the field size be $q_1 = 809$. Thus, after simple calculations, we see that the range of the masking random variable is $\{0,\ldots,39\}$. Let the user choose $x=[1,2]^t$. Thus, the queries sent to the two servers are given by \eqref{queries_no_weight}
and after the servers reply, the user reprocesses the received answers according to \eqref{pcr_ans_reprocess} to obtain
\begin{align}
    A^{[x]}(1) = \begin{bmatrix}
        \hat{A}_1^{[x]}(1)\\
        \hat{A}_2^{[x]}(1)
    \end{bmatrix} = \bm{M}_2\begin{bmatrix}
        365+\mu(1)\\
        I(1)+Z'(1)
    \end{bmatrix},
\end{align}
and
\begin{align}
    A^{[x]}(2) = \begin{bmatrix}
        \hat{A}_1^{[x]}(2)\\
        \hat{A}_2^{[x]}(2)
    \end{bmatrix} = \bm{M}_2\begin{bmatrix}
        325+\mu(2)\\
        I(2)+Z'(2)
    \end{bmatrix}.
\end{align}
It is clear that for any choice of $\mu(1)$, and $\mu(2)$, the user would be able to know that $y_2$ is closest to $x$.

Now, let us choose a larger field size, for example, $q_2 = 40009$ and apply the simple transform $T:a\rightarrow 10a$. Then, with simple calculations, we see that the range of the masking random variable is $\{0,\ldots,399\}$ which is larger than the previous one and still maintains the relative distance.

\paragraph{Leakage Analysis}
Similarly, on comparing the leakages of masking and baseline schemes, we show that the former is lower than the latter. The details of the proof are given in \cite{PCR_full_journal}.
% To compare the leakages of the masking and the baseline schemes, we proceed as 
% \begin{align}
%     I(y_{[M]} ; & d_i(x)+ \mu(i),i \in [M]| x) \nonumber\\
%     =& H(d_1(x) + \mu(1), \ldots, d_M(x) + \mu(M) | x) -H(d_i(x)+\mu(i), i \in [M]|y_{[M]}, x)\\
%     \leq& H(d_1(x) , \mu(1), \ldots, d_M(x) , \mu(M) | x) -H(\mu(i), i \in [M]|y_{[M]}, x)\\
%     =& H(d_1(x), \ldots, d_M(x) | x)  + H(\mu(1), \ldots, \mu(M) | x,d_1(x) , \ldots, d_M(x) )\nonumber \\
%     &-H(\mu(1),  \ldots, \mu(M)|y_{[M]}, x)\label{eq:mu_deterministic}\\
%     =&  H(d_1(x), \ldots, d_M(x) | x)  - H(d_1(x), \ldots, d_M(x) | x, y_{[M]}))\label{eq:mu_deterministic_reduced}\\
%     =& I(y_{[M]}; d_1(x), \ldots, d_M(x) | x),
% \end{align}
% where \eqref{eq:mu_deterministic_reduced} follows since the last two terms in  \eqref{eq:mu_deterministic} are equal, and $H(d_1(x), \ldots, d_M(x) | x, y_{[M]})$ is zero.

% \begin{remark}
%     The two schemes are not analytically comparable. This is because of the parameter $d_{\min}$ in the Mask-PCR scheme, which affects the amount of leakage, since the masking random variable $\mu$ depends on it. Moreover, $d_{\min}$ depends on $\mathcal{D}$, which is arbitrarily distributed. To resolve this, we numerically evaluate the leakage for different values of $d_{\min}$ and compare this with the Diff-PCR scheme (see Section \ref{expt:leakage_pcr}). 
% \end{remark}

\section{PCR WITH ACTIONABILITY}\label{private_pref}
In this section, we assume that the user assigns different \emph{weights} to their attributes based on their preference to change those attributes to attain a counterfactual. The more reluctant the user is to change a given attribute, the higher its weight is. In order to ensure that the user's preference is also private, the weight vector should be kept private from the servers. Let $w\in [L]^{d}$ denote the user's weight vector, where $L$ is a positive integer. Then, we have the following theorem.

\begin{theorem}\label{pcr_weighted_thm}
    There exist schemes derived from Baseline PCR, Diff-PCR and Mask-PCR, that can retrieve the exact closest counterfactual with user's actionability weights using $N = 3$ servers while keeping the user's sample and actionability weights hidden from each server. The communication costs of the respective schemes are:
    \begin{enumerate}
        \item Baseline PCR+: $6d+3M$ symbols of $\mathbb{F}_q$, $q>R^2 Ld$
         \item Diff PCR+: $6d+3(M-1)$ symbols of $\mathbb{F}_q$, $q> 2R^2Ld$
        \item Mask PCR+: $6d+3M$ symbols of $\mathbb{F}_q$, $q>R^2Ld$
    \end{enumerate}
    where $q$ is prime. 
\end{theorem}

To prove Theorem \ref{pcr_weighted_thm}, we describe the extended schemes next.{It is easy to verify that the PCR+ schemes guarantee actionability, at equal computational complexities as their PCR counterparts.}

\subsection{ Baseline PCR+}\label{baseline_weighted}
% \begin{theorem}\label{baseline_weighted_them}
%     There exists a scheme that can retrieve the exact closest counterfactual with user's actionability weights using $N = 3$ servers while keeping the user's sample and actionability weights hidden from each server. In addition, the total communication cost of this scheme is $3(2d +M)$ symbols of $\mathbb{F}_q$ where $q>R^2Ld$ is prime.
% \end{theorem}
% \vspace{-0.2cm}
 In this case, the required minimum field size is $q \geq R^2Ld$. Now, the user sends the query tuple $Q_n^{[x,w]}$ to server $n$, where
\begin{align}\label{queries_weighted}
    Q_n^{[x,w]}(1) = x+\alpha_n Z_1, \quad 
    Q_n^{[x,w]}(2) = w + \alpha_n Z_2,
\end{align}
and $Z_1$, and $Z_2$ are uniform independent random vectors in $\mathbb{F}_q^d$. Upon receiving the queries, the answers are generated as follows,
\begin{align}
   A_n^{[x,w]}(i) = &(y_i-Q_n^{[x,w]}(1))^t\diag(Q_n^{[x,w]}(2))\nonumber \\
   &\times (y_i-Q_n^{[x,w]}(1))+\alpha_n Z'_1(i) + \alpha_n^2 Z'_2(i)\\
     =& d_w(y_i,x) + \alpha_n \Big( (y_i-x)^t \bm{Z}(y_i-x)  \nonumber \\
     &-2(y_i-x)^t\bm{W} Z_1 +Z'_1\Big) +\alpha_n^2 \Big( Z_1^t \bm{W}Z_1 \nonumber \\
     &- 2(y_i-x)^t \bm{Z} Z_1 +Z'_2 \Big)+ \alpha_n^3 Z_1^t\bm{Z}Z_1,
\end{align}
where $\bm{W} = \diag(w)$, $\bm{Z} = \diag(Z_2)$, $Z'_1(i)$, and $Z'_2(i)$ are uniform random variables shared by the servers and used to invoke the one-time pad theorem.

Upon receiving the answers from $N=3$ servers, the user applies the following decoding approach,
\begin{align}
    \hat{A}_n^{[x,w]}(i) &= A_n^{[x,w]}(i) - \alpha_n^3 Z_1^t \bm{Z} Z_1,\\
    \hat{A}^{[x,w]}(i) & = \begin{bmatrix}
        \hat{A}_1^{[x,w]}(i),
        \hat{A}_2^{[x,w]}(i),
        \hat{A}_3^{[x,w]}(i)
    \end{bmatrix}^t \nonumber\\ &= \bm{M}_3\begin{bmatrix}
        d_{w}(y_i,x),
         I_1(i),
         I_2(i)
    \end{bmatrix}^t,
\end{align}
where
\begin{align}\label{interference_weighted}
 I_1(i) &= (y_i-x)^t \bm{Z}(y_i-x) -2Z_1^t \bm{W}(y_i-x) +Z'_1(i), \nonumber \\
 I_2(i) &= Z_1^t \bm{W}Z_1 - 2Z_1^t \bm{Z}(y_i-x) +Z'_2(i). 
\end{align}

The upload cost in this scheme is $6d$, while the download cost is $3M$.

\subsection{Diff-PCR+} 
% \begin{theorem}
%     There exists a scheme that provides lower leakage compared to the baseline PCR+ scheme with a communication cost $3(2d+M-1)$.
% \end{theorem}
In this case, the servers construct their answers so that the user decodes only the difference of distances, instead of the exact distances. This time, the field of operation is a prime $q> 2R^2Ld$, since, for all $i,j\in [M]$, the absolute difference $|d_w(y_i,x) - d_w(y_j,x)|=(y_i-x)^t \bm{W} (y_i-x)-(y_j-x)^t \bm{W} (y_j-x)$ is at most $R^2Ld$.
Therefore, in $\mathbb{F}_q$,
\begin{align}
    &d_w(y_i,x) - d_w(y_j,x)\in 
    \! \! \! \!
    &\begin{cases}
        \mathcal{H}, \! \! \! \! &\text{$y_j$ is closer to $x$},\\
        \mathcal{H}^c, \! \! \! \! & \text{$y_i$ is closer to $x$}, 
    \end{cases}
\end{align}
where $\mathcal{H}=[0:R^2Ld]$ and $\mathcal{H}^c = [R^2Ld+1:q-1]$. The user sends the same queries as in \eqref{queries_weighted}. Let the servers share two common randomness vectors ${Z}'_1=[Z'_1(1)\ldots Z'_1(M-1)]^t$ and ${Z}'_2=[Z'_2(1)\ldots Z'_2(M-1)]^t$ each of length $M-1$ where each entry is a uniform random variable from $\mathbb{F}_q$. Server $n\in \{1,2,3\}$ constructs the following answer for each $i\in [M-1]$
\begin{align}
    A_n^{[x,w]}(i)=&(y_i-Q_n^{[x,w]}(1))^t\text{diag}(Q_n^{[x,w]}(2))\nonumber \\
    &\times(y_i-Q_n^{[x,w]}(1))-(y_{i+1}-Q_n^{[x,w]}(1))^t\nonumber \\
    &\times \text{diag}(Q_n^{[x,w]}(2))(y_{i+1}-Q_n^{[x,w]}(1))\nonumber \\
    &+ \alpha_n Z_1'(i)+ \alpha_n^2 Z'_2(i)\\
 =& d_w(y_i,x) - d_{w}(y_{i+1},x)+\alpha_n I'_1(i) + \alpha_n^2 I'_2(i)
\end{align}
where  $I_1'(i)=(y_i-x)^t \bm{Z} (y_i-x) - (y_{i+1}-x)^t \bm{Z}(y_{i+1}-x)+2Z_1^t \bm{W}(y_i-y_{i+1})+Z'_1(i)$ and $I'_2(i)=2Z_1^t \bm{Z} (y_{i+1}-y_i)+Z'_2(i)$ are the interference terms.  The answers $A_n^{[x,w]}(i)$ of the three servers can be written as,
\begin{align}
    A^{[x,w]}(i)=\bm{M}_3
    \begin{bmatrix}
        d_w(y_i,x) - d_{w}(y_{i+1},x)\\
        I_1'(i)\\
        I_2'(i)
    \end{bmatrix},
\end{align}
where $A^{[x,w]}(i) = \begin{bmatrix}
        A_1^{[x,w]}(i),
        A_2^{[x,w]}(i),
        A_3^{[x,w]}(i)
    \end{bmatrix}^t$. Therefore, the user recovers the $M-1$ differences, while discarding the interference terms.

{Finding the Closest $y_i$ to $x$:} Now, with the $M-1$ differences, the user evaluates their counterfactual index $\theta^*$ following sequential comparisons similar to Algorithm \ref{alg_difference_distance}, the only difference being the range in line \ref{range_prox} is replaced by $\mathcal{H}^c$.

The upload cost in this scheme is $6d$, while the download cost is $3(M-1)$.

\subsection{Mask-PCR+}
% \begin{theorem}
%     There exists a scheme that satisfies the optimization problem defined in \eqref{opt_1} with weighted preferences being private from the servers. In addition, it provides less leakage compared to the baseline PCR+ scheme. The communication cost of this scheme is $3(2d+M)$.
% \end{theorem}

The queries are the same as in the previous section. Upon receiving the queries, each server applies the following on each sample $y_i$ and sends to the user
\begin{align}
    A_n^{[x,w]}(i) =& \big(y_i-Q_n^{[x,w]}(1)\big)^t\diag\big(Q_n^{[x,w]}(2)\big)  \nonumber \\&\times \big(y_i-Q_n^{[x,w]}(1)\big)+\mu(i) +\alpha_n Z'_1(i)\nonumber \\
    &+ \alpha_n^2 Z'_2(i)\\
    =& d_w(y_i,x)+ \mu(i) + \alpha_n I_1(i) \nonumber \\
    &+ \alpha_n^2 I_2(i) +\alpha_n^3 Z_1^t \bm{Z} Z_1,
\end{align}
where $\mu(i)$ is as defined in the previous section, $Z'_1(i)$, and $Z'_2(i)$ are uniform random variables and $I_1(i)$ and $I_2(i)$ are as given in \eqref{interference_weighted}. Using the answers of the $N=3$ servers and applying the decoding approach given in Section~\ref{baseline_weighted}, the user obtains the masked weighted distance corresponding to $y_i$ as follows:
\begin{align}
    & \hat{A}^{[x,w]}(i) =  \bm{M}_3\begin{bmatrix}
        d_w(y_i,x) +\mu(i),
         I_1(i),
         I_2(i)
    \end{bmatrix}^t.
\end{align}

To show that there is no need to change the range of the masking random variable $\mu$ with any weighting, we provide the following lemma, which is proven in \cite{PCR_full_journal}.

\begin{lemma}\label{lemma_range_fixed}
    The range of the random variable designed to mask the exact distance information in this case is the same as in the previous non-weighted case.
\end{lemma}

 The upload cost in this scheme is $6d$, and the download cost is $3M$.

For both Diff-PCR+ and Mask-PCR+, the proofs of lower leakage, in \eqref{eq:leakage_pcr}, compared to baseline PCR+ follow similarly to those of Diff-PCR and Mask-PCR.

\section{EXPERIMENTS} \label{sec_experiments}
\subsection{Accuracy-Quantization Trade-Off}\label{exp:accu_quant}
As detailed in Section \ref{userpriv1}, the proposed schemes operate on finite fields with size $q$. However, a real-world application might require some of the features to be real-valued. To circumvent this technicality, we may quantize each feature of the counterfactual instances as well as the user {sample} before applying the schemes {to align with our system model}. However, the quantization {of each feature to integers in $[0:R]$} will lead to a loss of some information that might result in an error in finding the closest counterfactual. The error can be made arbitrarily small by increasing the number of quantization levels, as we demonstrate in the following experiments. The trade-off is that a larger finite field $\mathbb{F}_q$ will be required for the scheme to work properly; e.g., for two implementations of Mask-PCR with numbers of quantization levels $R_1, R_2$ such that $R_1>R_2$, the minimum field sizes required are $q_1 \geq q_2$ because $q_1>R_1^2d$ and $q_2>R^2_2d$.

Moreover, in the case of Mask-PCR, we may lift the restriction $x\in\mathcal{D}_c$ off by empirically deciding the parameter $d_{\min}$ to be used such that the scheme works for most of the points in a sample set of queries. As pointed out in Section \ref{masking_equal_actionable}, a larger field size $q$ is needed to accommodate a larger value for $d_{\min}$.

We carry out our experiment using the Wine Quality Dataset \cite{wineDataset} to observe the accuracy-quantization trade-offs mentioned above. The dataset contains 4898 instances, each with 11 real-valued features. The original target variable ``Quality'' is categorical, taking integer values from 0 to 10. We convert the target into a binary variable by defining $\tau=\mathds{1}[``\text{Quality}"\geq 5]$. This gives us 3788 instances with $\tau=1$ and 183 instances with $\tau=0$ (which are the potential queries).

We consider all the instances with $\tau=1$ to be the accepted instances (potential counterfactual instances). Let this set be $\mathcal{S}_1$. The set $\mathcal{S}_0$ consists of all the instances with $\tau=0$ which are the potential queries. In each round, we pick a subset $\mathcal{D}$, with $|\mathcal{D}|=M$, of $\mathcal{S}_1$ as the database and a subset $\mathcal{S}_x$ of $\mathcal{S}_0$ as the queries. Then, we apply the retrieval schemes with varying parameters to obtain the nearest neighbor counterfactuals. The accuracy of the retrieved counterfactuals is computed as follows: Denote the counterfactual of the query $x$ retrieved using the scheme by $\hat{y}(x)$ and the actual (i.e., the exact nearest neighbor) counterfactual by $y(x)$. Note that the scheme provides the index of the counterfactual, hence, we can retrieve the exact unquantized counterfactual irrespective of the quantization applied. The accuracy of the scheme is  
\begin{align}
    \!\!\text{Accuracy} = \frac{1}{|\mathcal{S}_x|}\!\!\sum_{x_i\in\mathcal{S}_x} \!\!\mathds{1}\Big[||x_i-\hat{y}(x_i)||\leq ||x_i-y(x_i)||\Big].
\end{align}
We average the accuracy over several rounds with uniformly sampled $\mathcal{D}\subset\mathcal{S}_1$ and $\mathcal{S}_x\subset\mathcal{S}_0$. We set the database size $M$ to be 500 and the number of queries per round (i.e., $|\mathcal{S}_x|$) to be 50. We repeat the experiment for 100 rounds. 

\begin{figure}[t]
    \centering
    \includegraphics[width=0.8\columnwidth]{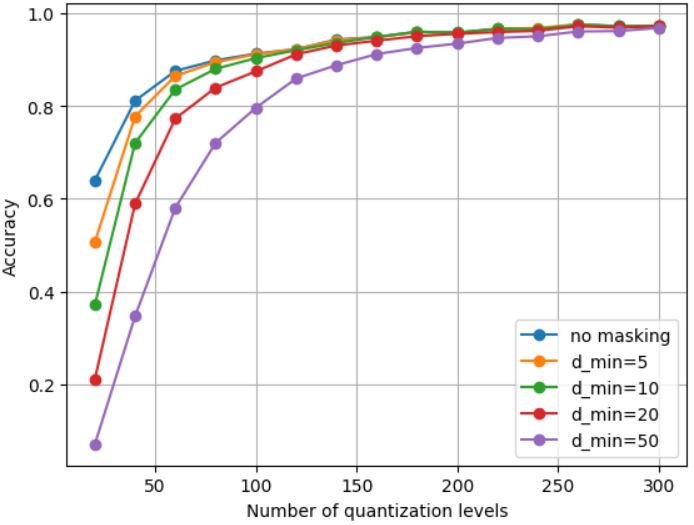}
    \caption{Accuracy-quantization trade-off.}
    \label{fig_accQuantizationTradeOff}
\end{figure}

Fig.~\ref{fig_accQuantizationTradeOff}, shows how the accuracy improves with an increasing number of quantization levels. Further, for a given number of quantization levels, the accuracy degrades as $d_{\min}$ increases.

\subsection{Database Leakage Results}\label{expt:leakage_pcr}
\subsubsection{Synthetic Dataset}
To observe the success of Diff-PCR and Mask-PCR in mitigating the leakage, we compute the exact leakage values over a synthetic dataset. We refer to \cite{PCR_full_journal} for leakage computation. We consider $R=3$ and $d=3$ with a database size of $M=3$. We assume the queries are equi-probable over the $[0:R]^d$ space. We further assume that given the query $x$, $y_1,\dots,y_M$ are equi-probable over $[0:R]^d\backslash \{x\}$. Table \ref{table_leakageValues} lists the computed leakage values under these assumptions.
\begin{table}[h!]
\centering
 \begin{tabular}{c c} 
 \hline
 Scheme & Leakage (to base $q=757$) \\ [0.5ex] 
 \hline
 Baseline PCR & 1.143 \\
 Diff-PCR & 0.940 \\
 Mask-PCR ($d_\text{min}=2$) & 0.911 \\
 Mask-PCR ($d_\text{min}=3$) & 0.777 \\
 \hline
 \end{tabular}
 \vspace{3mm}
 \caption{Leakage results -- synthetic data.}
 \label{table_leakageValues}
\end{table}

\subsubsection{COMPAS dataset}
Correctional Offender Management Profiling for Alternative Sanctions (COMPAS) dataset \cite{compas} includes 6172 instances and 20 features (after one-hot encoding the categorical features) about individuals, which are used to predict the likelihood of reoffending of a convicted criminal. The target variable $\tau$ is ``is\textunderscore recid''. Class-wise counts are 3182 and 2990 for $\tau=0$ and $\tau=1$, respectively. We normalize all the features to the interval $[0,1]$ and quantize to $R+1$ levels with $R=10$ during pre-processing, leaving 560 unique instances with $\tau=0$ and 717 unique instances with $\tau=1$. As our schemes are independent of the classifier, we use these labels as the classifications of the instances. In order to compute the leakage values, we assume the following:
\begin{enumerate}
    \item Queries are equi-probable over the $\tau=0$ (rejected) instances. 

    \item Since computing the histogram over all possible $M$-tuples (with $M=5$ in this particular experiment) is too computationally intensive, we sample $10^5$ of the possible permutations and evaluate the leakage over those. Accordingly, given the query $x$, the $M$-tuples $y=(y_1,\dots,y_M)$ are assumed to be equi-probable over a set of $10^5$ $M$-tuples that can be generated using the $\tau=1$ (accepted) instances. 
\end{enumerate}

Table \ref{table_leakageValuesCompas} presents the results. Similar to the values obtained with the synthetic dataset, we observe that Diff-PCR and Mask-PCR schemes reduce leakage compared to the baseline scheme.

\begin{table}[htb]
\centering
 \begin{tabular}{c c} 
 \hline
 Scheme & Leakage (to base $q=2003$) \\ [0.5ex] 
 \hline
 Baseline PCR & 1.51437 \\
 Diff-PCR & 1.51436 \\
 Mask-PCR ($d_\text{min}=2$) &  1.51427\\
 \hline
 \end{tabular}
 \vspace*{0.3cm}
 \caption{Leakage results -- COMPAS dataset.}
 \label{table_leakageValuesCompas}
\end{table}

% Moreover, we compute representative leakage values for the COMPAS \cite{compas} dataset as a real world example (see Table \ref{table_leakageValuesCompas}). Refer Appendix \ref{appendix_experiments} for more details on the implementation.

% References

 % has a nice set of citation styles and commands

\bibliographystyle{IEEEtran}
\bibliography{references.bib}

\end{document}